\def\Statusstring{}
\documentclass[12pt,letter]{article}

\usepackage{amsmath, amsthm, amssymb}
\usepackage{amssymb,epsfig}

\makeatletter

\gdef\@punct{.\ \ }  
\def\@sect#1#2#3#4#5#6[#7]#8{%
  \ifnum #2>\c@secnumdepth
     \def\@svsec{}
  \else
     \refstepcounter{#1}\edef\@svsec{%
     \ifnum #2>0{{\csname the#1\endcsname}}.\fi%
    \hskip .5em}
  \fi
  \@tempskipa #5\relax
  \ifdim \@tempskipa>\z@
     \begingroup #6\relax
       \@hangfrom{\hskip #3\relax\@svsec}{\interlinepenalty \@M #8\par}
     \endgroup
     \csname #1mark\endcsname{#7}
     \addcontentsline{toc}{#1}{\ifnum #2>\c@secnumdepth\else
          \protect\numberline{\csname the#1\endcsname}\fi#7}
  \else
     \def\@svsechd{#6\hskip #3\@svsec #8\@punct\csname
#1mark\endcsname{#7}
     \addcontentsline{toc}{#1}{\ifnum #2>\c@secnumdepth \else
          \protect\numberline{\csname the#1\endcsname}\fi#7}}
  \fi
  \@xsect{#5}}

\def\@ssect#1#2#3#4#5{\@tempskipa #3\relax
  \ifdim \@tempskipa>\z@
     \begingroup #4\@hangfrom{\hskip #1}{\interlinepenalty \@M
#5\par}\endgroup
  \else \def\@svsechd{#4\hskip #1\relax #5\@punct}\fi
  \@xsect{#3}}

\def\qed{\hskip 3pt \hbox{\vrule width4pt depth2pt height6pt}}

\newtheorem{Lemma}{Lemma}

\newtheorem{Theorem}[Lemma]{Theorem}

\newtheorem{Corollary}[Lemma]{Corollary}

\begin{document}

\title{Minimal resolving sets for the hypercube}

\author{Ashwin Ganesan%
  \thanks{Department of Mathematics, Amrita School of Engineering, Amrita Vishwa Vidyapeetham, Ettimadai, Coimbatore - 641~105, India.
  Email: \texttt{ashwin.ganesan@gmail.com, g\_ashwin@cb.amrita.edu}. }}

\date{}

\maketitle

\vspace{-4.5cm}
\begin{flushright}
  \texttt{\Statusstring}\\[1cm]
\end{flushright}
\vspace{+2.5cm}

\begin{abstract}
\noindent  For a given undirected graph $G$, an \emph{ordered} subset $S = \{s_1,s_2,\ldots,s_k\} \subseteq V$ of vertices is a resolving set for the graph if the vertices of the graph are distinguishable by their vector of distances to the vertices in $S$.  While a superset of any resolving set is always a resolving set, a proper subset of a resolving set is not necessarily a resolving set, and we are interested in determining resolving sets that are minimal or that are minimum (of minimal cardinality).  Let $Q^n$ denote the $n$-dimensional hypercube with vertex set $\{0,1\}^n$.  In Erd\"os and Renyi \cite{Erdos:Renyi:1963} it was shown that a particular set of $n$ vertices forms a resolving set for the hypercube.  The main purpose of this note is to prove that a proper subset of that set of size $n-1$ is also a resolving set for the hypercube for all $n \ge 5$ and that this proper subset is a minimal resolving set.

\end{abstract}




\section{Introduction}

Let $G=(V,E)$ be a simple, undirected graph.  An ordered subset $S = \{s_1,s_2,\ldots,s_k\} \subseteq V$ of vertices is a resolving set for the graph if the vertices of the graph are distinguishable by their vector of distances to the vertices in $S$.  In other words, if $d(v,S)$ denotes the vector $(d(v,s_1),d(v,s_2),\ldots,d(v,s_k))$ of distances in the graph from $v$ to the elements of $S$, then $S$ is a resolving set for $G$ if and only if $d(u,S) \ne d(v,S)$ whenever $u \ne v$.  In this manner, the vertices in $S$ `resolve' or are able to distinguish between the vertices of the graph using information on the distances to these vertices.  Note that this definition immediately implies that a superset of any resolving set is again a resolving set, but a proper subset of a resolving set is not necessarily a resolving set.  A well-studied problem has been to obtain resolving sets which are minimum (i.e. of minimal size) or which are minimal (i.e. which have no proper subsets that are resolving sets).    The concept of resolving sets arises both in puzzles such as the coin weighing problem and strategies for the Mastermind game, as well as practical applications such as chemical compounds and drug discovery \cite{Chartrand:etal:2000}, network discovery and verification \cite{Beerliova:etal:2005}, and robot navigation.  Thus, it is of interest to have descriptions of resolving sets which are minimal; see \cite{Caceres:etal:2007} for some recent results.

The $n$-dimensional hypercube $Q^n$ is the graph whose vertices are the $2^n$ 0-1 sequences of length $n$, with two vertices being adjacent whenever the corresponding two sequences differ in exactly one coordinate.  Resolving sets for hypercubes were studied in Erd\"os and Renyi \cite{Erdos:Renyi:1963}, Lindestr\"om, and many other papers (see \cite{Caceres:etal:2007} for some recent results), and this problem still remains open, though some asymptotic bounds are known.  A lower bound due to Erd\"os and Renyi and an upper bound due to Lindestr\"om imply that the minimum size of a resolving set of $Q^n$ asymptotically approaches $2n / \log n$.  The problem of determining minimum resolving sets for the $n$-dimensional hypercube for finite values of $n$ is still open.  Resolving sets for more general graphs were first introduced in Harary and Melter \cite{Harary:Melter:1976}; the minimum size of a resolving set of a graph is also called the metric dimension of the graph.

\bigskip \noindent \textbf{Notation: }  We use the following notation for hypercubes and level sets of the poset of subsets of an $n$-element set \cite{Bollobas:1986}. Let $X$ denote the $n$-element set $\{1,2,\ldots,n\}$.  With a slight abuse of notation, we use the same symbol $v \in V(Q^n)$ for a vertex of the hypercube to represent both a binary string of length $n$ as well as a subset of $\{1,2,\ldots,n\}$.  The operation $+$ denotes binary addition.  Thus, for $n=5$, we can write $x = 11001 = \{1,2,5\}$, $y=10100=\{1,3\} \in V$, and $|x+y| =|01101| = |\{2,3,5\}|=3$, where the operation of binary addition of two strings is equivalent to that of taking the symmetric difference of the two corresponding sets.  Also, $e_i \in V$ is the binary string that has a 1 in the $i$-th coordinate and a 0 in the remaining coordinates; thus, $e_2 = 010\ldots0=\{2\}$.  The level set $X^{(k)}$ denotes the set of all subsets of $X$ of size $k$.   For $v \in V$ and an ordered subset $S = \{s_1,s_2,\ldots,s_k\} \subseteq V$, we define the distance vector $d(v,S) := (d(v,s_1),d(v,s_2),\ldots,d(v,s_k))$, where the distance $d(x,y)$ between two vertices $x$ and $y$ in $Q^n$ is just the number of coordinates where the two sequences differ, which equals the size of their symmetric difference.  Thus, $S$ is a resolving set for $Q^n$ if $u \neq v$ implies $d(u,S) \neq d(v,S)$.  For $x \in V$ and $S \subseteq V$, $S+x$ denotes the (ordered) set obtained by adding $x$ to each element of $S$.

\subsection{Prior work and our results}

In Erd\"os and Renyi \cite{Erdos:Renyi:1963}, the set of 4 vertices $\{ \{1,2,3,4,5\},$ $\{1,2,3\},$ $ \{2,4\}, \{2,3,5\} \}$ was shown to be a resolving set for $Q^5$; they provided the results of an exhaustive computation of the distance vectors for each of the 32 vertices of $Q^5$ to this set and observed that they are all distinct.  More recently, Caceres et al \cite{Caceres:etal:2007} provide a particular set of 4 vertices in $Q^5$ which they say can be verified to be a resolving set by a laborious calculation.   Furthermore, Caceres et al \cite{Caceres:etal:2007} also provide some upper bounds on the minimum size of resolving sets of $Q^n$ for $n \le 15$ obtained through exhaustive computer searches. The problem of obtaining minimum resolving sets for the hypercube or even just determining the minimum size of a resolving set is still open for small values of $n$.  We show here that for the general case $n \ge 5$, a particular set of $n-1$ vertices which is simple to describe is a resolving set and minimal.  Unlike the case of the previous two papers, our proof does not rely on exhaustive computation.

In Erd\"os and Renyi \cite{Erdos:Renyi:1963} it was shown that the set of $n$ vertices $\{11\ldots1, 011\ldots1,$ $1011\ldots1, \ldots, 111\ldots101\}$ is a resolving set for $Q^n$. Our results imply that a proper subset of this set of size $n-1$ is also a resolving set for the $Q^n$. In particular, we show that the $n-1$ vertices $\{011\ldots1, 1011\ldots1,$ $ \ldots, 111\ldots101\}$ are a minimal resolving set for $Q^n$.  However, let us also mention that the main result of Erd\"os and Renyi \cite{Erdos:Renyi:1963} was not to provide a resolving set of size $n$ for $Q^n$ which we improve upon here but an asymptotic lower bound on the minimum size of a resolving set for $Q^n$; we just provide an alternate proof for $n=5$ that does not rely on an exhaustive computation and we strengthen one of their results for finite $n$.

\section{Main Results}

\bigskip \noindent It will be useful to apply the following lemma, which follows immediately from the symmetry properties of the hypercube.

\begin{Lemma} \label{lemma1} Let $x \in V$. $S$ is a resolving set for $Q^n$ if and only $S+x$ is a resolving set for $Q^n$.
\end{Lemma}

\noindent \emph{Proof: }  $S$ is not a resolving set if and only if there exist distinct vertices $u,v \in V$ such that $d(u,S) = d(v,S)$.  Observe that $d(u,S) = d(u+x, S+x)$.  Hence $S$ is not a resolving set if and only if there exist distinct vertices $u' = u+x$ and $v'=v+x$ such that $d(u',S+x)=d(v',S+x)$.  This is the case if and only if $S+x$ is not a resolving set.
\hfill\qed.

\bigskip \noindent Thus, we can always assume that $\phi$ is an element of any resolving set of $Q^n$.  Note that the distance between a vertex $u$ of the hypercube that lies in the $k$-th level set $X^{(k)}$ of the poset and the vertex $\phi$ is $k$.  Thus, if $S$ is a resolving set of $Q^n$ and $\phi \in S$, then to show $d(u,S) \ne d(v,S)$ whenever $u \ne v$, it suffices to show that $d(u,S) \ne d(v,S)$
whenever $u$ and $v$ are distinct vertices that lie in the same level set of the hypercube.

\begin{Lemma}
The set $\{\phi, \{1\}, \{2\}\}$ is a minimum resolving set for $Q^3$.  The set $\{\phi, \{2\}, \{3\}, \{4\}\}$ is a minimum resolving set for $Q^4$.
\end{Lemma}

\noindent \emph{Proof sketch: }
The proof is straightforward and relies on a case by case analysis.  To show that the sets given in the assertion are resolving sets, we use the fact that $\phi$ belongs to this set, and hence, it suffices to check that for any distinct vertices $u,v$ in the same level set of the poset, the distance vectors $d(u,S)$ and $d(v,S)$ are distinct.

To show that $Q^3$ does not have a resolving set of size 2, we let $S=\{\phi,v\}$.  If $S$ is a resolving set, then $|v|$ equals 1,2 or 3, and for each case, it is seen that one can find two distinct vertices having the same distance vector to $S$, resulting in a contradiction.  Hence, the resolving set given in the assertion can be seen to be of minimum size.

For $n=4$, to show that there does not exist a resolving set of $Q^4$ of size less than 4, suppose $S=\{\phi,x,y\}$ is a resolving set for $Q^4$.  Suppose $x \in X^{(k)}, y \in X^{(\ell)}$.  By doing a case-by-case analysis for the different $(k,\ell)$ values, we find that for each case, there exists some $u \ne v$ such that $d(u,S)=d(v,S)$.~\hfill\qed.

We now prove the main result.

\begin{Theorem}  \label{maintheorem}
For $n \ge 5$, $S = \{e_2,e_3,\ldots,e_n\} = \{ \{2\}, \{3\}, \ldots, \{n\} \}$ is a minimal resolving set for $Q^n$.
\end{Theorem}

\noindent \emph{Proof: }  Let $S = \{e_2,e_3,\ldots,e_n\} = \{ \{2\}, \{3\}, \ldots, \{n\} \}$.  For any $i \ge 2$, it is clear that any proper subset $S-\{e_i\}$ is not a resolving set because the two vertices $e_1$ and $e_i$ will have the same distance vector to $S - \{e_i\}$.  By Lemma~\ref{lemma1}, it now suffices to prove that $S' := S+e_2$ is a resolving set.  Note that $S'$ is the set of vertices $\{\phi, \{2,3\}, \{2,4\}, \ldots, \{2,n\}\}$.   Let $x,y$ be any two distinct vertices of $Q^n$.  We now show that $d(x,S') \neq d(y,S')$.  Since $\phi \in S'$, it suffices to show this lack of equality for just the case where $x$ and $y$ are in the same level set $X^{(k)} \subseteq V$.  For the rest of this proof, we assume $x \neq y, ~x,y \in X^{(k)}$ and $1 \le k \le n-1$.  For each value of $k$, we exhibit an $s \in S'-\{\phi\}$ such that $d(x,s) \neq d(y,s)$.

Suppose $k=1$.  If $x = e_1$ and $y = e_i \subseteq \{2,\ldots,n\}$, then we let $s$ be any element in $S'-\{\phi\}$ that contains $i$.  Observe that $d(x,s)=3$, but $d(y,s) =1$.  If $x=e_2$ and $y = e_i \subseteq \{3,\ldots,n\}$, we take $s$ to be any element (in $S'-\{\phi\}$, as always) that does not contain $i$. If $1,2 \not\in x,y$, pick $s$ to be any element that contains $x$ but not $y$.  Thus, for all $x,y \in X^{(1)}$, there exists an element in the resolving set $S'$ which distinguishes $x$ from $y$ based on the distance vector to $S'$.

Suppose $k=2$.  We can assume $x,y \in X^{(2)} - S'$, since if either $x$ or $y$ is in $S'$, one of the components in exactly one of the two distance vectors will be zero.  We examine three different choices for $x$ and $y$.  (i) First, suppose $2 \in x$.  Then $x=\{1,2\}$ and $2 \not\in y$.  If $1 \in y$, then pick $s$ to be any element that is disjoint from $y$, and observe that $d(x,s)$ is smaller than $d(y,s)$ because $x$ and $s$ overlap but $y$ and $s$ are disjoint.  If $1 \not\in y$, then, since $n \ge 5$, there exists an $s$ disjoint from $y$ satisfying our requirement.  (ii) Second, suppose $2\in y$.  This case is resolved just like the previous case. (iii)  Suppose $2 \not\in x,y$.  Then, there exists an $i \in \{3\ldots, n\}$ such that $i \in x, i \not\in y$.  Pick $s = \{2,i\}$.

Finally, suppose $k \ge 3$. Then $x$ and $y$ each have a nonempty intersection with $\{3,\ldots,n\}$.  If $x \cap \{3,\ldots,n\} = y \cap \{3,\ldots,n\}$, then we may assume that $1 \in x, 2 \in y, 1 \not\in y, 2 \not\in x$, and in this case we can take $s$ to be any element. Thus, we may now suppose that there exists an $i \in \{3,\ldots,n\}$ such that either $i \in x, i \not\in y$ or $i \in y, i \not\in x$.  Without loss of generality, assume the former.  Consider two cases:  (i) Suppose $2 \in x$.  Then we can take $s=\{2,i\}$, and observe that $d(x,s) = |x|-|s|$ because $s \subset x$, whereas $d(y,s) > |y|-|s| = |x|-|s|$ because $i \not\in y$.  (ii)  Now suppose $2 \not\in x$.  We consider two subcases.  The first subcase is $2 \not\in y$.  Then we can take $s=\{2,i\}$, and observe that $d(y,s) \ge d(x,s)+2$.  The second subcase is $2 \in y$.  Let $j$ be any element in $y \cap \{3,\ldots,n\}$, and take $s=\{2,j\}$.  Observe that $s \subset y$, but $s \not\subset x$.~\hfill\qed

We now show that a proper subset of the resolving set given in \cite{Erdos:Renyi:1963} is also a resolving set.

\begin{Corollary}
Let $n \ge 5$. The set $\{011\ldots1, 1011\ldots1,$ $ \ldots, 111\ldots101\}$ is a minimal resolving set for $Q^n$.
\end{Corollary}

\noindent \emph{Proof: } The same proof used to show that $S=\{e_2,e_3,\ldots,e_n\}$ is a minimal resolving set can also be used, mutatis mutandis, to show that $\hat{S}=\{e_1,e_2,\ldots,e_{n-1}\}$ is a minimal resolving set. (Equivalently, any permutation of the $n$ coordinates induces a vertex automorphism of the hypercube graph, and automorphisms preserve distances.)   By Lemma~\ref{lemma1}, $\hat{S}+11\ldots1$ is a minimal resolving set.~\hfill\qed

\section{Further remarks}

We showed that the resolving set of size $n$ provided in Erd\"os and Renyi \cite{Erdos:Renyi:1963} was not minimal, and we showed that a proper subset of their set of size $n-1$ is a resolving set and is minimal.  We now explain how some other resolving sets of size $n-1$ can be constructed by combining the resolving set of size 4 for $Q^5$ given in Erd\"os and Renyi \cite{Erdos:Renyi:1963} (or in Caceres et al \cite{Caceres:etal:2007}) with a result on the resolving set of products of graphs given in Chartrand et al \cite{Chartrand:etal:2000}.

Let $G=H \times K_2$ be the cartesian product of $H$ and $K_2$, which consists of two copies of $H$, say $H_1$ and $H_2$, where $H_1$ has resolving set $\{w_1,\ldots,w_k\}$ and $H_2$ has resolving set $\{u_1,\ldots,u_k\}$, and the distance in $G$ between $w_i$ and $u_i$ is 1.  Then, it is shown in \cite{Chartrand:etal:2000} that $\{w_1,\ldots,w_k,u_1\}$ is a resolving set for $G$, and hence, that the metric dimension of $G$ is at most 1 more than the metric dimension of $H$.  Since $Q^n$ is the product of $Q^{n-1}$ and $K_2$, it follows that one can use the resolving sets of size 4 for $Q^5$ given in \cite{Erdos:Renyi:1963} or \cite{Chartrand:etal:2000} to successively construct resolving sets of size $n-1$ for $Q^n$.  Thus, since $\{\phi,e_2\}$ is a resolving set for $Q^2$, it follows that $\{\phi,e_2,e_3,\ldots,e_n\}$ is a resolving set of size $n$ for $Q^n$.  The smallest value of $n$ for which the metric dimension of $Q^n$ is less than $n$ is $n=5$.  Hence, by using a minimum resolving set for $Q^5$, it is possible to construct resolving sets of size $n-1$ for $Q^n$.  Nonetheless, the methods in the literature that were used to prove that certain sets for $Q^5$ are resolving sets are based on brute force calculations or on an exhaustive computer search, and so our proof is an alternative method and also provides new resolving sets.

Note that by the symmetry of the hypercube, Theorem~\ref{maintheorem} characterizes all the minimal resolving sets of the hypercube that lie entirely in the first level set $X^{(1)}$ or in the other level set $X^{(n-1)}$ of the poset of subsets of $\{1,2,\ldots,n\}$.  It can be shown that each of the remaining level sets $X^{(k)}, 2 \le k \le n-2$ also forms a resolving set.  A natural further problem is to characterize the \emph{minimal} resolving sets that lie entirely in any of the remaining level sets $X^{(k)}$ of the poset or that satisfy certain other prescribed conditions.

\bibliographystyle{plain}
\bibliography{refscomb}

\end{document}